\documentstyle[12pt]{article}
\def\sq{\hbox{\vrule\vbox{\hrule\phantom{o}\hrule}\vrule}}

\begin{document}
\begin{titlepage}

Preprint \hfill   \hbox{\bf SB/F/98-253}

\hrule 
\vskip 1.5cm 
\centerline{\bf  BF models, Duality and
Bosonization on higher genus surfaces} 
\vskip 1cm

\centerline{A.Restuccia and J.Stephany \footnote{Regular Associate of
the Abdus Salam International Centre for Theoretical Physics}}
\vskip 4mm
\begin {itemize}
\item{\ }{\it  Universidad Sim\'on
Bol\'{\i}var, Departamento de F\'{\i}sica, Apartado Postal 89000,
Caracas 1080-A, Venezuela.}
\item{\ }{\it \ \ \ e-mail: arestu @usb.ve, stephany@usb.ve}
\end{itemize}
\vskip 1cm

{\bf Abstract} \vskip .5cm

The generating functional of two dimensional $BF$ field theories 
coupled to fermionic fields and conserved currents is computed in 
the general case when the base manifold is a genus g compact 
Riemann surface. The lagrangian density $L=dB{\wedge}A$
is written in terms of a globally defined 1-form $A$ and a multi-valued 
scalar field $B$. Consistency conditions on the periods of $dB$  
have to be imposed. It is shown that there exist  a non-trivial 
dependence of the generating functional on the topological restrictions
imposed to  $B$. In particular if the periods of the
$B$ field are constrained to take values $4\pi n$, with $n$ any
integer, then the partition function is independent of the
chosen spin structure and may be written as a sum over all the
spin structures associated to the fermions even when one started
with a fixed spin structure. These results are then applied to the
functional bosonization of fermionic fields on higher genus
surfaces. A bosonized form of the partition function which takes
care of the chosen spin structure is obtained.

\vskip 1cm
\hrule
\bigskip
\centerline{\bf UNIVERSIDAD SIMON BOLIVAR}
\vskip .5cm
\vfill
\end{titlepage}

\section{Introduction}
 In this paper we compute the generating functional of $BF$
\cite{H} topological systems coupled to fermions on a two
dimensional compact manifold of arbitrary genus and apply the
result to discuss the bosonization \cite{Boz} of the fermions. This
work has two different motivations. Firstly, it has been observed
\cite{LR}, \cite{LRS}, that the allowed world hyper-surfaces
described by classical sources (p-branes) coupled to a $BF$ theory
are subject to restrictions of topological nature . One is then
led to ask the question of how this effect is translated to the
quantum theory.  The second aspect which motivates this
work concerns the relation between topological models and
duality transformations. For a large class of systems, duality
transformations have been devised along the lines of the T-duality
transformation in the sigma models \cite{TB}. The method consists
essentially in a two step elimination of one field in terms of it
dual variable. First one introduces an auxiliary gauge field
constrained at the beginning to have zero curvature. This allows
to decouple the original variable from the currents and then one
may perform the remaining integral in the quadratic
approximation. At this point  the connection to the
$BF$ \cite{H} theory appears, since to impose the zero
curvature condition into the functional integral one may introduce
the partition function of a $BF$ topological model. When the
duality transformation is applied to the generating functional of 
free fermionic fields on genus
zero manifolds, \cite{BQ}, \cite{FS}, it leads to the bosonized
representation of the theory \cite{Boz}. 
In the operatorial approach bosonization in two
\cite{Boz} and three dimensions \cite{M3}, \cite{MS}, \cite{KK}
has been related to the construction of dual soliton operators
\cite{SM} in bosonic theories and this give an additional meaning
to the duality transformation discussed in \cite{BQ} and
\cite{FS}. At the intermediate step after introducing the gauge
field one is dealing with a $BF$ theory coupled to fermions which
is the subject of this paper. This point of view has been also
extended to higher dimensions \cite{Boz3}.

Over topologically trivial manifolds, the procedure described
above allows in some cases to determine exact equivalences  between
fields theories. When the base manifold has  genus
$g$ one has to take care of the global definition of the geometrical
objects appearing in the formulation \cite{HV},\cite{JS}. The
global aspects introduced by the auxiliary fields in the path
integral has been explicitly tested for example in the purely
bosonic self-dual vectorial model in $3-D$. This model is known to
be locally equivalent to the topologically massive model
\cite{DJT} and in fact can be viewed as a gauge fixed version of
it \cite{GRS}. Nevertheless it has been shown that in
topologically non trivial manifolds this equivalence has to be
reinterpreted \cite{JS}, \cite{AR}, \cite{PS} since the partition
function of the topologically massive model has an additional
factor of topological origin.  When matter fields are
included, the coupling with the topological field theory may be
related to self interaction terms for the fermions \cite{GMD}.

In the specific case of the fermionic models there are other
reasons to explore the consequences of defining the system on
higher genus surfaces. Even in genus zero surfaces, when coupled
to gauge fields with non-trivial topological properties, the
fermions show dynamical effects. The most notorious of these is
the non vanishing of fermion condensates {\cite{DL}, \cite{CLS},
\cite{CS}} due to the contributions of the instantons associated
in four dimensions to the resolution of the $U(1)$ problem
\cite{tH}. The vanishing of such condensates in the topological
trivial case is enforced by gauge symmetry. The non vanishing
result in the most general situation may be traced in the
functional approach to an explicit contribution of the zero modes
of the fermionic fields{\cite{DL}. In higher genus surfaces one
expects a more rich structure in the gauge field sector, but also
further complications
 are introduced in the
duality transformation when gauge fields of non-vanishing
topological index have to be considered. For this reason in this
article  we exclude this possibility.

This paper is organized as follows. In section 2 we review some
useful concepts and notation and discuss the result of the
computation of the fermionic determinant in genus $g$ manifolds.
In section 3 we compute the generating functional of a particular
$BF$ topological theory coupled to fermions. This model which as
we will see later appears naturally in the bosonization of
fermions in higher genus surfaces is described in terms of a
1-form $A$ globally defined and a multivalued $B$ field. 
Since $dB$ should remain univalued one has to impose restrictions
over the periods of $dB$. When these periods are chosen to be
integral multiples of $4\pi$ the partition function is shown to be
a sum over all spin structures even if one starts with a fixed
spin structure. This is an interesting result which in particular
implies that the partition function is independent of the spin
structure originally chosen.  In section 4 we discuss the
bosonization \cite{Boz} of fermionic fields on higher genus Riemann
surfaces. Here again, bosonization may be understood as a duality
transformation \cite{FS},\cite{BQ} between the fermionic current
and the Hodge dual of the field intensity tensor of a vector
field.} A careful treatment of the global aspects in the
formulation leads naturally to a bosonized effective action in terms
of a multi-valued 0-form.

\section{ The fermionic determinant on higher genus compact Riemann
surfaces}

 We will consider $BF$ models coupled to fermionic fields over
higher genus Riemann surfaces. In order to compute its partition
function one needs the explicit formula for the fermionic
determinant in the case of zero curvature gauge potentials $A$ on
trivial $U(1)$ line bundles. This determinant was computed in
references \cite{LAG}, \cite{FP}.  To express the result, let
us introduce some notation concerning the properties of the
manifold and the fields. We take ${ a_{i}}$ and ${b_{j}}$ to be a
basis of homology of closed curves over $\Sigma$, a compact
Riemann surface of genus $g$. The set of curves ${ a_{i}}$ and
${b_{j}}$ will be denoted by ${\cal C}^I$. If one deforms
continuously the fermionic field along the curves of the basis,
after returning to the original point the fermionic field may change
sign or not. A spin structure over $\Sigma$ is determined by a
combination of one of this two possibilities for each of the curves of
the basis. The gauge potentials $A$ on a trivial $U(1)$ bundle are
characterized by the vanishing of the Chern class

\begin{equation}
\int_{\Sigma}F(A)=\int_{\Sigma}dA=0 .
\end{equation}
The index of the corresponding Dirac operator is then zero and
consequently there are no zero modes in the fermionic sector.

The potential may be decomposed into its exact, co-exact
and harmonic parts:
\begin{equation}
\label{decomp}
A=ds+^*dp+A_{h} .
\end{equation}
The  harmonic part of the field is expressed in terms of a base of
real harmonic forms $\alpha_{i}$ and $\beta_{i}$, $i,j=1,\cdots,g$ as
follows,
\begin{equation}
\label{harm}
A_{h}=2\pi\sum_{i}^g (u_{i}\alpha_{i}-v_{i}\beta_{i}) .
\end{equation}
The real harmonic basis $\alpha_{i}$ and $\beta_{j}$ is constructed
from two normalized holomorphic basis $\omega_{j}$ and
$\hat{\omega}_{j} $, $ j=1,\cdots, g$
\begin{eqnarray}
\int_{a_{i}}\omega_{j}=\delta_{ij}, {\hskip .3cm}
\int_{a_{i}}\hat{\omega}_{j}=\hat{\Omega}_{ij}\nonumber\\
\int_{b_{i}}\omega_{j}={\Omega}_{ij}, {\hskip .3cm}
\int_{b_{i}}\hat{\omega}_{j}=\delta_{ij}, 
\end{eqnarray}
where $\Omega$ is the period matrix. In terms of $\omega_{j}$ and
$\hat{\omega}_{j}$, $\alpha_{i}$ and $\beta_{j}$ are given by
 \begin{eqnarray}
 \beta_{j}=\frac{1}{2i}(\omega_{k}-\bar{\omega_{k}})[Im
 \Omega]^{-1}_{kj} \nonumber\\
 \alpha_{i}=\frac{1}{2i}(\hat{\omega_{k}}-\bar{\hat{\omega_{k}}})[Im
 \hat{\Omega]}^{-1}_{ki} .
 \end{eqnarray}
The imaginary part of the period matrix $Im\Omega$ is always an
invertible matrix.  Let us now consider a fermionic field
defined over a genus $g$ compact Riemann surface with a definite
but arbitrary spin structure. The spin structure is fixed by
specifying two $g$-dimensional vectors $\epsilon_{i}$ and
$\kappa_{j}$ with components $0$ or $\frac{1}{2}$ so that the
periodicities of the fermions about the cycles $a_{i}$ and $b_{j}$
are respectively $exp(2{\pi}i\epsilon_{i})$ and
$exp(-2{\pi}i\kappa_{j})$.  The partition function which
defines the fermionic determinant is

\begin{equation}
\label{zhgenus}
\hat{Z}_f[A,\epsilon,\kappa]=\int{\cal D}\psi{\cal D}\bar{\psi}
e^{\int_\Sigma d^2x \sqrt{g}
\bar\psi(-i{D\!\!\!/} + {A\!\!/})\psi } =det[-i{D\!\!\!/} + {A\!\!/}]
,
\end{equation}
where we take ${D\!\!\!/}$ to be the covariant derivative for
the fermions.

The fermionic determinant in this situation may be obtained from the
results in \cite{LAG},\cite{FP} and is given by ,
\begin{eqnarray}
\label{detg}
\hat{Z}_f[A,\epsilon,\kappa] = &\\
e^{-\frac{1}{2\pi}\int_{\Sigma}{F(A)\frac{1}{\Delta_{0}}{\ ^*}F(A)}}&
\left[\frac{det Im \Omega
Vol(\Sigma)}{det'\Delta_{0}}\right]^{\frac{1}{2}}
\left|\theta \left[ \begin{array}{c}
 u + \epsilon \\
  v + \kappa
 \end{array} \right](0|\Omega)\right|^2 .\nonumber
\end{eqnarray}
Here $\Delta_{0}$ is the Laplacian operator acting on $0$-forms
and $Vol(\Sigma)$ is the area of the Riemann surface. The third
factor is a $\theta$-function given by, 
\begin {equation}
\label{theta} \theta \left[ \begin{array}{c}
        u  \\
        v
\end{array}\right](0|\Omega)= \sum_{n \in {\cal
Z}^g}exp[i{\pi}(n+u){\Omega}(n+u) + i2{\pi}(n+u)v] .
\end{equation}

 We note that in (\ref{detg}) the first factor depends only on
the co-exact component of the gauge field. This contribution
corresponds to the result for genus zero surfaces \cite{HRS} which
is usually written in the form,

\begin{equation}
\label{det}
\frac{det(-i{\partial\!\!/} + {A\!\!/})}{det(-i{\partial\!\!/})}=
e^{-\frac{1}{2\pi}\int d^2x \sqrt{g}
A_\mu(\delta^{\mu \nu} - \frac{\partial^\mu\partial^\nu}{\sq})A_\nu}.
\end{equation}
The other two factors in (\ref{detg}) give the Dirac determinant
for a purely harmonic potential $A_{h}$ of the form (\ref{harm}).
The result (\ref{detg}) has been used to investigate the Schwinger
model in higher genus surfaces \cite{FF}.

 Finally by summing over all spin structures, we may also
define,
\begin{equation}
\label{sumspin}
\hat{Z}_f[A]=\sum_{\epsilon,\kappa}\hat{Z}_f[A,\epsilon,\kappa]
\end{equation}
which will play a role in what follows.

\section{Two dimensional BF theories coupled to fermions}

 In this section we will compute the generating functional for
a particular $BF$ system coupled to fermions over a genus $g$
Riemann surface. The action functional of a $BF$ theory is
written in terms of a connection $A$ and a field $B$ which may be
interpreted as a Lagrange multiplier which enforces the $A$ field
to have zero curvature \cite{H}. In its usual form it is given by,
 \begin{equation}
S_{BF}= \int_{\Sigma}{dA\wedge B}
 \end{equation}
Here the $0$-form $B$ and $dA$ are defined globally on the
manifold $\Sigma$. The connection $A$ may be allowed to have
transitions over $\Sigma$. The computation of the partition
function of this system was discussed in \cite{BT}. The off-shell
BRST charge was computed in \cite{CR}. We will consider a modification
of this system  which appears naturally in the
context of bosonization. We consider the action,
\begin{equation}
S_{BF}^{mod}= \int_{\Sigma}{dB\wedge A}
\end{equation}
which may be different from the action above, for trivial bundles,
only over higher genus surfaces. The one forms $A$ and $dB$ have
to be globally defined but $B$ may be multi-valued. Due to the non
trivial topological structure of the manifold, one may distinguish
three cases in the definition of the generating functional. One
may consider the following conditions on the periods of $dB$:
\begin{eqnarray}
\label{Cir0}
\oint_{C^I}dB=0\\
\label{Cir2n}
\oint_{C^I}dB=2{\pi}m^I \\
\label{Cir4n}
\oint_{C^I}dB=4{\pi}m^I .
\end{eqnarray}
The first case is the usual $BF$ model.
In the second and third cases we consider the summation on all the
values of $m^I$ in the functional integral (\ref{BF}). Each of the
choices defines a different model.

 The generating functional of these systems coupled to
conserved currents $j$ and $J$ is given in all the cases by
\begin{equation}
\label{BF}
Z[j,J,\epsilon,\kappa]=\sum_{m^I}\int{\cal D}C{\cal D}A{\cal D}B{\cal
D}\psi{\cal
D}\bar{\psi}  e^{-S_{eff}} ,
\end{equation}
\begin{equation}
S_{eff}=\int d^2x \sqrt{g}
[{\bar\psi}(i{D\!\!\!/} - {A\!\!/} -
{j\!\!/})\psi + L_{g}]+\int_{\Sigma}(\frac{i}{2\pi}dB{\wedge}A
-{\ ^*J}{\wedge}A)
\end{equation}
where $L_{g}$ includes the gauge fixing term and the contributions
of the auxiliary fields (ghosts fields and Lagrange multipliers)
and ${\cal D}C$ stands for the integration measure in those
fields. The sum in $m^I$ is included to stress the fact that we
are summing over the $B$ field configurations which satisfy either
(\ref{Cir0}), (\ref{Cir2n}) or (\ref{Cir4n}).  The spin
connection is fixed and identified by the $g$-dimensional vectors
$\epsilon_{i}$ and $\kappa_{j}$ .

 The functional
integration on the Lagrange multiplier $B$ of course provides the
factor $\delta(F(A))$ in the measure of the generating functional
but as we will see presently , the additional summation over the
periods gives rise to a factor which constraints also the periods
of $A$. Let us see how this works. Suppose for example that we
compute the generating functional (\ref{BF}) summing over the $B$
field configurations which fulfills (\ref{Cir4n}). Given two
different configurations of $B$, say $B_{1}$ and $B_{2}$,
satisfying this condition we have
\begin{equation}
 B_{2}-B_{1}=b ,
\end{equation}
where $b$ is univalued over $\Sigma$. In general we may then write,
 \begin{equation}
\label{b}
 B=B_{m^I}+b ,
\end{equation}
with $B_{m^I}$ a specific configuration satisfying
(\ref{Cir4n}) with a set of values $m^I$.
The functional integration on the multi-valued $B$ field has been
expressed as an integration on
the univalued function $b$ and a sum over all possible choices $m^I$.
Consider now the $BF$ action in the sector defined by one of such
choices. We have
\begin{eqnarray}
\label{GBF}
 \int_{\Sigma}{dB\wedge A}= \int_{\Sigma}{d(B_{m^I}A)}+
 \int_{\Sigma}{(-B_{m^I})\wedge dA}+ \int_{\Sigma}{(-b)\wedge dA} .
\end{eqnarray}
The generating functional becomes,
\begin{equation}
\label{GBF2}
Z[j,J,\epsilon,\kappa]=\sum_{m^I}{\int{\cal D}A{\cal D}b{\cal
D}C{\cal D}\psi{\cal D}\bar{\psi}e^{- {\cal S}_{eff}}}
\end{equation}
\begin{eqnarray}
 {\cal
S}_{eff}=\int_{\Sigma}d^2x{\sqrt{g}[\bar{\psi}(i{D\!\!/} -
{j\!\!/}-{A\!\!/})\psi+L_{g}]} \nonumber\\+{\frac{i}{2\pi}}
\int_\Sigma{[d(B_{m^I}A)
- (B_{m^I}+b){\wedge}dA] -\int_{\Sigma}{\ ^*J}{\wedge}A}\nonumber
\end{eqnarray}

At this point, one recovers the factor $\delta(dA)$
in the measure of the generating
functional by performing the functional integral in $b$ and in the
ghost fields introduced to guarantee the $BRST$ invariance of the
effective action. In particular this makes the second term in
(\ref{GBF}) to vanish and to disappear also from (\ref{GBF2}).

To evaluate the remaining functional integral, consider a
triangulation of
$\Sigma$ in terms of elementary domains $U_{i}$, $i\in [1,\ldots,N]$.
Since $\Sigma$ is compact the triangulation exists and the covering
is
provided by a finite number of elementary domains. Let $A^{i}$ and
$B_{m^I}^i$ be the restrictions of the fields to the domain $U_{i}$.
Then, in
the functional space projected by $\delta(dA)$, we have
\begin{eqnarray}
\left[\int_{\Sigma}{dB\wedge
A}\right]_{\mid_{dA=0}}=\int_{\Sigma}{d(B_{m^I}A)} = \sum_{i=1}^N
{\int_{U_{i}}{d(B_{m^I}^iA^{i})}} \nonumber\\  =
\sum_{U_{i}\cap U_{j}\neq \emptyset }{\int_{U_{i}\cap U_{j}}
{(B_{m^I}^i-B_{m^I}^j)A}} = \sum_{U_{i}\cap U_{j}\neq \emptyset}
{4\pi m^{(ij)}\int_{U_{i}\cap U_{j}}{A}}
\end{eqnarray}
where $ m^{(ij)}$ are integers. Using again that the connection is
flat we finally get,
\begin{equation}
 e^{\frac{i}{2\pi}\int_{\Sigma}{dB\wedge A}}= e^{i\sum_{I}2
m^{I}\oint_{C^I}A}.
 \end{equation}
 Here one recognizes the coefficients of the Fourier expansion of a
delta function with period $\pi$. Upon summing over all $m^I$ the
total contribution is  \begin{equation}
 \label{delpe}
        \delta(F(A))\sum_{I}\delta(\oint_{{\cal C}^I}A - {\pi}n^I) .
 \end{equation}
 When the $B$ field in (\ref{BF}) is taken to satisfy (\ref{Cir0})
only
 the factor $\delta(F(A))$ appears. We will not discuss this case
furthermore. When the $B$ field is taken to satisfy
 (\ref{Cir2n}), the second delta function has period $2\pi$ and the
factor turns out to be
 \begin{equation}
\label{delpe2}
        \delta(F(A))\sum_{I}\delta(\oint_{{\cal C}^I}A - 2{\pi}n^I) .
 \end{equation}
  Let us see now how the conditions (\ref{delpe}) or
(\ref{delpe2}) enter in the complete evaluation of (\ref{BF}).
Using the decomposition (\ref{decomp}) for the $A$ field, the
 factor $\delta(dA)$ in the measure of (\ref{BF}) allows the
integration of the co-exact part of $A$ and we are left with the task
of determining which are the configurations of $A_h$ that
contribute.  It is now straightforward to show that the delta
functions in (\ref{delpe}) (or respectively (\ref{delpe2}))
constrain the values of the coefficients in the expansion
(\ref{harm}) of $A_h$ to be half-integers (or integers).
To continue we use this fact and perform the functional
integration in the fermions. Defining $u^0$ and $v^0$to
 be the coefficients in the expansion of the harmonic part of $j$,
\begin{equation}
j_{h} = 2\pi\sum_{i}^g (u_{i}^0\alpha_{i}-v_{i}^0\beta_{i}) .
 \end{equation}
 we obtain:
\begin{eqnarray}
 \label{j,J}
 Z[j,J,\epsilon,\kappa]=&&\nonumber \\
 \sum_{u,v}\left[\frac{det Im \Omega Vol(\Sigma)}
 {det'\Delta_{0}}\right]^{\frac{1}{2}}
\left|\theta \left[\begin{array}{c}
 u + u^0 + \epsilon \\
  v + v^0 + \kappa
 \end{array}\right](0|\Omega)\right|^2 e^{\int_{\Sigma}
 (^*J{\wedge}A_{h}-\frac{1}{2\pi}dj\frac{1}{\Delta_{0}}^*dj)} ,
\end{eqnarray}
 The sum in (\ref{j,J}) is over the allowed values of $u$ and $v$
which as we already said are all the integers or all the
half-integers depending which case we are considering. From here
on  we have to
distinguish between the two cases.

Let us take first the case when the $B$ field satisfies
(\ref{Cir2n}).  Then in (\ref{j,J}) we have a sum over the
$g$-tuples with integral entries which we label by $m$ and $l$. The
factor with the theta function in (\ref{j,J}) takes the form,
\begin {equation} \label{sqm} \left|\theta \left[\begin{array}{c}
 u^0 + m + \epsilon \\
 v^0 + l +\kappa
 \end{array}\right](0|\Omega)\right|^2
\end {equation}
It is straightforward to see from (\ref{theta}) that this becomes
independent of $l$ due to the square norm that we are taking.
Moreover (\ref{sqm}) also is independent of $m$, since in
(\ref{theta}) one may redefine $n+m=n'$ and one still will have 
summation in all $n'$. We can then factorize the contribution of the
harmonic part of the field to the partition function in the form,
\begin{equation}
\label{ssdep}
Z_{2n}[j,J,\epsilon,\kappa] =  \hat{Z}_f[j,\epsilon,\kappa]
\sum_{m,l} e^{\int_{\Sigma}{\ ^*J}{\wedge}A_{h}}
\end{equation}
with $\hat{Z}_f[j,\epsilon,\kappa]$ given by (\ref{detg})
\begin{equation}
\hat{Z}_f[j,\epsilon,\kappa]= \left[\frac{det Im \Omega Vol(\Sigma)}
{det'\Delta_{0}}\right]^{\frac{1}{2}}
\left|\theta\left[
\begin{array}{c}
 u^0 + \epsilon \\
 v^0 + \kappa
\end{array}
\right](0|\Omega)\right|^2
e^{-\frac{1}{2\pi}\int_{\Sigma}
dj\frac{1}{\Delta_{0}}{\ ^*d}j}
\end {equation}
Note that the external current $J$  only couples to the harmonic part
of the vector field. When $J$ is zero we obtain,
\begin{equation}
\label{J0}
Z_{2n}[j,0,\epsilon ,\kappa]={\cal N}\hat{Z}_f[j,\epsilon,\kappa]
\end{equation}
with ${\cal N}$ a  constant which measures the volume of the harmonic
space. This factor is expected from the original expression
(\ref{BF}) since in that case the volume of the zero modes factorizes
from the functional integral.

Consider now the situation when (\ref{Cir4n}) holds. We have
instead of (\ref{sqm}) the expression,
 \begin {equation}
\label{sqm2} \left|\theta \left[\begin{array}{c}
 u^0 + \frac{m}{2} + \epsilon \\
 v^0 + \frac{l}{2} +\kappa
 \end{array}\right](0|\Omega)\right|^2
\end {equation}
where $\frac{m}{2}$ and $\frac{l}{2}$ are the half integer periods of
$A$. We consider the following decomposition,
\begin{eqnarray}
\frac{m}{2}=& m'+\eta, \nonumber \\
\frac{l}{2}=&l'+\mu
\end{eqnarray}
where $m'$ and $l'$ are integer numbers while $\eta$ and $\mu$ are
$g$-tuples with components $0$ or $\frac {1}{2}$. Summation in all $m$
and $l$ is equivalent to summation in all $(\eta,\mu)$ and all
$(m',l')$.  The summation in the integers may be handled
 as before. Then the summation in the half integers $(\eta,\mu)$ may
be reinterpreted as a sum over all spin structures (weighted by a
factor which depends on $J$). When $J$ is zero we have,
\begin{equation}
\label{z4n}
Z_{4n}[j,0,\epsilon,\kappa]= {\cal
N}\sum_{\epsilon',\kappa'}\hat{Z}[j,\epsilon',\kappa']={\cal N}Z_f[j,]
\end{equation}
The factor ${\cal N}$ here  gives the same measure of the space of
harmonic 1-forms with integral periods as in (\ref{J0}). We started
with a fixed spin structure, however the final result corresponds to
the partition function of spinor fields with summation
n in all spin structures. In particular it shows that
$Z_{4n}[j,0,\epsilon,\kappa]$ is independent of the spin structure
(i.e of $\epsilon$ and $\kappa$).

\section{Bosonization in higher genus surfaces}

As an application we use the results of the previous section
to discuss the bosonization of fermions over higher genus compact
Riemann surfaces. The equations (\ref{ssdep}) and (\ref{z4n})
already establish the relation between the partition function of
the fermions and the partition function of the $BF$ model. In this
section we obtain this result using the
the constructive approach of \cite{BQ} and \cite{FS}.

Let us begin with a quick review of the situation in the
topologically
trivial case. Consider the  generating functional of a fermion field
coupled to a conserved current $j$,
\begin{equation}
Z_f[j]=\int{\cal D}\psi{\cal D}\bar{\psi} e^{\int d^2x \sqrt{g}
\bar\psi(-i{\partial\!\!/} +
{j\!\!/})\psi }.
\end{equation}
We suppose here that the current $j$ has a topological index zero.
In two dimensions, on a genus zero surface, this fermion
determinant is explicitly known \cite{HRS} and given by
(\ref{det}). The duality-bosonization transformation allows to
express this result in terms of a bosonic field. To construct this
transformation one begins observing that the system has a global
$U(1)$ gauge invariance.  Then \cite{FS},\cite{BQ},\cite{TB}
one makes a change of variables with the functional form of 
a local gauge transformation and identify the spurious contributions
which appear in the action as coupling terms with a gauge field of
zero curvature. The adequate change of variables in this case is,

\begin{equation}
\label{change}
\psi(x)\longrightarrow e^{i\Lambda(x)}\psi(x)
\end{equation}
where $\Lambda(x)$ is an arbitrary parameter with local dependence on
$x$. The
fermionic generating functional turns out to be,

\begin{equation}
Z_j[j]= {\cal K}\int{\cal D}\psi{\cal D}\bar{\psi}e^{\int d^2x
\sqrt{g} \bar\psi(-i{\partial\!\!/} +
{j\!\!/} + {\partial\!\!/\Lambda})\psi } .
\end{equation}
where ${\cal K}$ is the Jacobian of the transformation (which in this
case is a  non-relevant constant).
This can be re-interpreted as the partition function of a
model consisting of a flat connection $A_{\mu}$ coupled to the
fermions in the particular gauge where,
 \begin{equation}
\label{Amu}
 A_\mu =  \partial_\mu\Lambda .
 \end{equation}
 The zero curvature condition on $A_{\mu}$  implies, of course,
that the connection is locally a pure gauge. Since the vanishing
of $\ ^*F(A )= \epsilon^{\mu \nu}F_{\mu\nu}(A)$ implies that of
$F_{\mu\nu}(A)$, one introduces the 1-form connection restricted
by the condition

\begin{equation}
 \ ^*F(A)  =  \epsilon^{\mu \nu}F_{\mu\nu}(A) = 0 .
\end{equation}
 After imposing this constraint in the functional integral one
gets,
\begin{equation}
Z_j[j]=\int{\cal D}A{\cal D}\psi{\cal D}\bar{\psi} {\frac{\delta
\ ^*(F(A))}
{Vol({\cal G}_A)}} e^{\int d^2x \sqrt{g} \bar{\psi}(-i{\partial\!\!/}
+
{j\!\!/} + {A\!\!/})\psi } ,
\end{equation}
where ${\cal G}_A$ is the gauge group of $A$. Now one introduces  a
Lagrange
multiplier $B$ to raise the $\delta(F)$ to the exponential but has
to take into account that since there are infinitely many solutions
of the equation $\ ^*F(A)=0$,   the functional
$\delta(\ ^*F)$ has to be defined with some care. It is properly
defined, \cite{JS} in terms of the generating
functional of a $BF$ topological field theory  \cite{CR}.  Using the
$BRST$ invariance as a guide to guarantee that the functional
integral remains well defined we get,
\begin{equation}
Z_f[j]=\int{\cal D}A{\cal D}B{\cal D}\psi{\cal D}\bar{\psi}{\cal D}C
e^{\int
d^2x  \sqrt{g}(\bar{\psi}(-i{\partial\!\!/} +
{j\!\!/} + {A\!\!/})\psi -{\frac{i}{2\pi}}\epsilon^{{\mu
\nu}}\partial_{\mu}B A_{\nu} + L_{g})} ,
\end{equation}
 where again ${\cal D}C$ stands for the measure of the ghosts
and auxiliary fields and $L_{g}$ for the contributions of those
fields plus the gauge fixing term to the Lagrangian. The appearance
of the $BF$ effective action should be expected since the factor
which comes from the exterior derivative in $\delta (\ ^*F(A))$ may be
expressed as a function of the Ray-Singer torsion an hence related
to the $BF$ effective action \cite{BT}. In two dimensions the
Ray-Singer torsion turns out to be equal to one.

To complete the bosonization of the generating functional one
 makes a shift $A + j \rightarrow A$. The fermionic field
remains coupled only to the new $A$ field. Then one  uses the
result (\ref{det}) for the fermionic determinant,  chooses an
adequate gauge
fixing condition which allow to make the quadratic functional
integral
in $A$ and ends up with,

\begin{equation}
Z_f[j]=Z[0]{\cal N}\int{\cal D}B e^{-\int d^2x \sqrt{g}
({\frac{1}{4}}\partial_{\mu}B\partial_{\mu}B
-{\frac{i}{2\pi}}\epsilon_{{\mu
\nu}}\partial_{\mu}B j_{\nu})} ,
\end{equation}
where ${\cal N}$ is the factor which appears after the
quadratic integral on $A$ has been performed. This is the
bosonized effective action. The external current $j$ appears in
this expression coupled to the topological current of the Lagrange
multiplier $B$.

Let us now turn to the general case on an arbitrary genus $g$,
compact Riemann surface. On the light of (\ref{z4n}) we start
with,
\begin{equation}
\label{zfer}
 \hat{Z}_f[j]=\sum_{\epsilon_{i},\kappa_{j}}\int{\cal D}\psi{\cal
D}\bar{\psi} e^{\int d^2x \sqrt{g} \bar\psi(-i{\partial\!\!/} +
{j\!\!/})\psi }
\end{equation}
Instead of using directly (\ref{z4n}) let us argue how one can
adapt the discussion presented for the genus zero surfaces and recover the
$BF$ partition function in a constructive way.  Let us
introduce the change of variables (\ref{change}). In order to have
a uniform change of variables in the functional integral,
$\Lambda(x)$ must satisfy

\begin{equation}
\label{Circulation}
\oint_{{\cal C}^{I}}d\Lambda = \pi n^{I}
\end{equation}
where $n^{I}$ are integers. If all the $n^{I}$ are
even the change of variables does not change the spin structure
that we have defined over $\Sigma$. Otherwise we change from one
to another spin structure but since we are summing over all of
them this is not a problem here. We get again,

\begin{equation}
\hat{Z}_f[j]=\sum_{\epsilon_{i},\kappa_{j}}\int{\cal D}\psi{\cal
D}\bar{\psi}e^{\int_{\Sigma} d^2x \sqrt{g} \bar\psi(-i{\partial\!\!/}
+
{j\!\!/} + {\partial\!\!/\Lambda})\psi } .
\end{equation}

In this case we also wish to rewrite this in terms of a
globally defined flat connection $A$. For two dimensional surfaces
this means that $A$ should be a flat connection over a trivial
$U(1)$ line bundle. To achieve consistency with
(\ref{Circulation}) we have to impose that
\begin{equation}
\label{Circulation2}
G(A)=\oint_{{\cal C}^{I}}A  =  \pi n^{I} .
\end{equation}
This is exactly the condition forced by(\ref{delpe}) and in
fact its appearance at this point provided the original motivation
to the discussion presented in the previous section. Things now
follow smoothly. First, in order to introduce $A$ satisfying
(\ref{Circulation2}) in the functional integral one extends the
functional integral to the space of  connections and introduces
factors $\delta(F(A))$ and $\delta(G(A))$ in the
measure. We get,

\begin{equation}
\label{ZD}
\hat{Z}_f[j]=\int{\cal D}A{\cal D}\psi{\cal D}\bar{\psi}
{\frac{\delta (F(A))
\delta (G(A))}{Vol({\cal G}_A)}} e^{\int_{\Sigma} d^2x \sqrt{g}
\bar{\psi}(-i{\partial\!\!/} +
{j\!\!/} + {A\!\!/})\psi } ,
\end{equation}
where ${\cal G}_A$ is the group of allowed gauge transformations
of $A$, that is of those gauge transformations with an uniform
gauge function.

Now we want to raise the $\delta$ functions to the exponential. From
our results  of the previous section, the right way to do that
 is to take  a multi-valued
Lagrange multiplier B over $\Sigma$ satisfying
\begin{equation}
\label{Circulation3}
\oint_{{\cal C}^{I}}dB = 4\pi m^{I}
\end{equation}
and to integrate over the functional space of $B$ with all possible
$m^{I}$. In order to have a well defined functional integral,
the measure has to be defined in terms of precisely the $BF$
topological field
theory we considered previously. We then recover (\ref{z4n})
\begin{equation}
\label{Bmulti}
\hat{Z}_f[j]= Z_{4n}[j,0,\epsilon,\kappa]=
\sum_{m^I}\int{\cal D}C{\cal D}A{\cal D}B{\cal
D}\psi{\cal
D}\bar{\psi}  e^{-S_{eff}} ,
\end{equation}
\begin{equation}
S_{eff}=\int d^2x \sqrt{g}
[{\bar\psi}(i{D\!\!\!/} - {A\!\!/} -
{j\!\!/})\psi + L_{g}] +\frac{i}{2\pi}\int_{\Sigma}(dB{\wedge}A)
\end{equation}
Here as we discussed earlier the result does not depend on the spin
structure $(\epsilon,\kappa)$.
To obtain the bosonized representation of (\ref{zfer}) we now  choose
the gauge fixing and ghost terms in (\ref{Bmulti}) and perform the
fermionic integral. We can work more generally with $J\neq 0$ and use (\ref{BF}).
Making first a shift
\begin{equation}
\tilde{A}=A+j
\end{equation}
in (\ref{BF}), taking the gauge condition
\begin{equation}
{\ ^*d}{\ ^*}{\tilde{A}}=0
\end{equation}
and performing the fermionic integral we have,
\begin{eqnarray}
\label{Lorentz}
Z_{4n}[j,J,\epsilon,\kappa]=
{[det'{\Delta}_0]^{\frac{1}{2}}}
\left[{det Im \Omega Vol(\Sigma)}\right]^{\frac{1}{2}}
\left|\theta \left[ \begin{array}{c}
u + \epsilon \\
v + \kappa
\end{array} \right](0|\Omega)\right|^2\nonumber\\ 
\times \ \ \sum_{m^I}\int {\cal D}B{\cal
D}\tilde{A}e^{-S[\tilde{A},B]}\\
S[\tilde{A},B]=
{\frac{1}{2\pi}}\int_{\Sigma}[d\tilde{A}{\frac{1}{\Delta_0}}{\
^*d\tilde{A}} + idB\wedge(\tilde{A}-j)+\nonumber\\
{\frac{1}{2}}d{\ ^*\tilde{A}}\wedge{\ ^*d\ ^*\tilde{A}} -2\pi{\
^*J}\wedge(\tilde{A}-j)]
\end{eqnarray}
where a factor $[det'{\Delta}_0]$ arises from the
integration on the ghost and anti ghost fields. The arguments $u$
and $v$ in the theta function are the coefficients in the
expansion of $\tilde{A}_h$ and are not restricted until now. To
write out our final expression we introduce the decomposition
(\ref{decomp}) for $\tilde{A}$ ,
$$ 
\tilde{A}=d\tilde{s}+{\ ^*}d\tilde{p}+\tilde{A}_{h} 
$$ 
and observe that, {\it i)}
Integration in $\tilde{s}$ contributes with a factor
$(det'\Delta_0)^{-1}$ {\it ii)} Integration in $\tilde{p}$ and the jacobian of the transformation contribute a factor
$(det'\Delta_0)^{\frac{1}{2}}$ and a term in the action of the form, 
$$
S(B,J) = -{\frac{1}{2\pi}}\int_\Sigma (dB+2{\pi}i{\
^*J})_{exact}\wedge{\ ^*} (dB+2{\pi}i{\ ^*J})_{exact} 
$$ 
since only the exact part of $(dB+2{\pi}i{\ ^*j})$ couples with
$\tilde{p}$ {\it iii)} One is left with the integration in
$\tilde{A}_{h}$. Using the decomposition (\ref{b}), for the the
field B one may show again that the summation over the periods of
$B$ leads to the half integral periodicity conditions in
$\tilde{A}_{h}$. The integral in $\tilde{A}_{h}$ is then a
summation over the half-integral periods. We finally obtain,
\begin{equation}
Z_{4n}[j,J,\epsilon,\kappa]=
\sum_{l,m}\int{\cal D}b e^{-S[\tilde{A}_{h},b]}
\left[det Im \Omega Vol(\Sigma)\right]^{\frac{1}{2}}
\left|\theta \left[ \begin{array}{c}
\frac{l}{2} \\
\frac{m}{2}
\end{array}\right](0|\Omega)\right|^2
\end{equation}
where the contribution of the spin structure is
included in the argument of the $\theta$ function and we define,
\begin {eqnarray}
S[\tilde{A}_{h},b]=
{\frac{1}{2\pi}}\int_\Sigma(db+2{\pi}i{\ ^*J}_{exact})
\wedge{\ ^*}(db+2{\pi}i{\ ^*J}_{exact})\\
-i(db+2{\pi}{\ ^*J}_{exact}){\wedge}j + \int_\Sigma{\
^*J}{\wedge}\tilde{A}_{h}\nonumber
\end{eqnarray}
with $A_{h}$ given by (\ref{harm}) restricted to half-integers periods.
When $J$ is zero this gives the bosonized expression for the
fermionic partition function in higher genus surfaces.  A
similar expression for the partition function over a single spin
structure may be obtained straightforwardly, following the same
lines, starting from (\ref{ssdep}). The results presented in this
paper allow to encode the information concerning the spin
structure of the manifold in terms of the topological properties
of the fields of a $BF$ model. They show also the non-trivial way
in which the bosonization rules are generalized to higher genus
surfaces.

\section{Acknowledgments}
J.S thanks S.A.Dias, R.L.G.P do Amaral and H.Christiansen for
useful discussions and the Abdus Salam International Centre for
Theoretical Physics, specially the High Energy Group for hospitality
during the completion of this work.

\end{document}